\documentclass[a4paper,10pt,twoside]{cpc-hepnp}

\usepackage{multicol}
\usepackage{graphicx}
\usepackage{booktabs}
\usepackage{amssymb,bm,mathrsfs,bbm,amscd}
\usepackage[tbtags]{amsmath}
\usepackage{lastpage}
\begin{document}
\fancyhead[c]{\small Chinese Physics C~~~Vol. 39, No. 10 (2015) 106201}
\fancyfoot[C]{\small 106201-\thepage}

\footnotetext[0]{Received 3 February 2015, Revised 10 April 2015}

\title{A data analysis method for isochronous mass spectrometry using two time-of-flight detectors at CSRe \thanks{Supported by the 973 Program of China (2013CB834401), National Nature Science Foundation of China (U1232208, U1432125, 11205205,11035007) and the Helmholtz-CAS Joint Research Group (Group No. HCJRG-108).}}

\author{%
      XU Xing$^{1,2,3;1)}$\email{xuxing@impcas.ac.cn}%
\quad WANG Meng$^{1;2)}$\email{wangm@impcas.ac.cn}%
\quad SHUAI Peng$^{4}$
\quad CHEN Rui-jiu$^{1}$
\quad YAN Xin-liang$^{1}$
\quad ZHANG Yu-hu$^{1}$\\
\quad YUAN You-jin$^{1}$
\quad XU Hu-shan$^{1}$
\quad ZHOU Xiao-hong$^{1}$
\quad Yuri A. Litvinov$^{1,5}$
\quad Sergey Litvinov$^{5}$
\quad TU Xiao-lin$^{1,5}$\\
\quad CHEN Xiang-cheng$^{1,2}$
\quad FU Chao-yi$^{1,2}$
\quad GE Wen-wen$^{1,2}$
\quad GE Zhuang$^{1,2}$
\quad HU Xue-jing$^{1,2}$\\
\quad HUANG Wen-jia$^{1,2,3}$
\quad LIU Da-wei$^{1,2}$
\quad XING Yuan-ming$^{1,2}$
\quad ZENG Qi$^{4}$
\quad ZHANG Wei$^{1,2}$
}
\maketitle

\address{%
$^1$ Key Laboratory of High Precision Nuclear Spectroscopy and Center for Nuclear Matter Science, Institute of Modern Physics, Chinese Academy of Sciences, Lanzhou 730000, China\\
$^2$ University of Chinese Academy of Sciences, Beijing, 100049, China\\
$^3$ CSNSM-IN2P3-CNRS, Universit\'{e} de Paris Sud, F-91405 Orsay, France\\
$^4$ Research Center for Hadron Physics, National Laboratory of Heavy Ion Accelerator Facility in Lanzhou and University of Science and Technology of China, Hefei 230026, China \\
$^5$ GSI Helmholtzzentrum f\"{u}r Schwerionenforschung, Planckstra{\ss}e 1, 64291 Darmstadt, Germany\\
}

\begin{abstract}
  The concept of isochronous mass spectrometry (IMS) applying two time-of-flight (TOF) detectors originated many years ago at GSI. However, the corresponding method for data analysis has never been discussed in detail. Recently, two TOF detectors have been installed at CSRe and the new working mode of the ring is under test. In this paper, a data analysis method for this mode is introduced and tested with a series of simulations. The results show that the new IMS method can significantly improve mass resolving power via the additional velocity information of stored ions. This improvement is especially important for nuclides with Lorentz factor $\gamma$-value far away from the transition point $\gamma _t$ of the storage ring CSRe.
\end{abstract}



\footnotetext[0]{\hspace*{-3mm}\raisebox{0.3ex}{$\scriptstyle\copyright$}2013
Chinese Physical Society and the Institute of High Energy Physics
of the Chinese Academy of Sciences and the Institute
of Modern Physics of the Chinese Academy of Sciences and IOP Publishing Ltd}%

\begin{multicols}{2}

\section{Introduction}
Nuclear masses play an essential role in our understanding of nuclear structure far from the valley of stability and origin of elements in the cosmos \cite{Klaus}. However, precision mass measurements of exotic nuclei are strongly restricted by their low production rates and short half-lives. Isochronous mass spectrometry (IMS) based on a heavy-ion storage ring, which was pioneered in the early 2000s at ESR-GSI in Darmstadat \cite{HauNIMA_IMS,HauHPFI_IMS} and then also successfully performed at CSRe-IMP in Lanzhou \cite{TUXLPRL_11,ZHANGPRL_12,YANAPJL_13,SHUAIPLB_14,XUHS}, has been proven to be a powerful tool for mass measurements of short-lived exotic nuclei. In particular, since the revolution times of stored ions can be measured with a time-of-flight~(TOF) detector \cite{MEIBO}, nuclei with half-lives as short as a few tens of microseconds can be measured. In an IMS experiment, the exotic nuclei are produced and separated with a fragment separator and then stored in a storage ring.

The revolution times $(T)$ of ions stored in a storage ring are a function of their mass-over-charge ratios $(m/q)$ and velocities $(v)$ in the first order approximation \cite{Franzke,RadonRPL} as follows:
\begin{eqnarray}
\begin{aligned}
\frac{\Delta{T}}{T}=-\frac{\Delta{f}}{f}
& \approx \frac{1}{\gamma_t^2} \frac{\Delta(m/q)}{m/q}-(1-\frac{\gamma^2}{\gamma_t^2})\frac{\Delta{v}}{v} \\
& \approx \alpha_p \frac{\Delta(m/q)}{m/q}-(1-\alpha_p \gamma^2)\frac{\Delta{v}}{v}, \label{eq1}
\end{aligned}
\end{eqnarray}
where ${\alpha}_p$ is the momentum compaction factor, which is defined as $\alpha_p=\frac{1}{\gamma_t^2}\equiv\frac{\Delta{C}/C}{\Delta{B\rho}/{B\rho}}$~connecting the relative variation of the close orbit length of stored ions and the relative variation of their magnetic rigidity. $\gamma_t$ is the so-called transition point of the ring \cite{Franzke}.

According to Eq.~(\ref{eq1}), it is clear that the nuclear masses of stored ions
can be determined from their revolution times when the second term on the right hand side is negligible \cite{Franzke}. One technique is to reduce the velocity spread of the stored ions to the level of $\sigma_v/v\approx 10^{-7}$ \cite{GeisselRPL} by means of an electron cooling device. This is so-called Schottky Mass Spectrometry (SMS) \cite{Franzke}, where the revolution times are measured non-destructively with Schottky pick-ups. Another technique is to tune the velocity of the target ion so as to fulfil the ${\gamma}={\gamma_t}$ condition, where $\gamma$ is the Lorentz factor of that ion. This is so-called Isochronous Mass Spectrometry \cite{HauNIMA_IMS,HauHPFI_IMS}, where the ion's revolution time does not depend on its velocity in the first order approximation.

For a given species of ions, the relative difference of its revolution time $\frac{\Delta{T}}{T}$ is determined by its velocity difference as described in Eq.(\ref{eq2}) \cite{Franzke}:
\begin{equation}
\frac{\Delta{T}}{T}=-(1-\frac{\gamma^2}{\gamma_t^2})\frac{\Delta{v}}{v}=-{\eta}\frac{\Delta{p}}{p}\label{eq2},
\end{equation}
where $\Delta p/p =\gamma ^2 \Delta v/v$ is the momentum difference, and ${\eta}=\frac{1}{\gamma^2}-{\frac{1}{\gamma_t^2}}$ is the so-called phase slip factor \cite{Franzke}. $\Delta p/p$ is limited by  the $B\rho$ acceptance and is almost the same for all ion species, while the $\eta$-values of different ion species vary significantly depending on their mean velocities or their $m/q$ ratios.

From Eq.({\ref{eq1}}) and ({\ref{eq2}}), the mass resolving power can be defined as \cite{Klaus}:
\begin{equation}
R=\frac{m}{2.355\ \sigma _m} \approx \frac{1}{\gamma_t^2}\frac{T}{2.355\ \sigma _T}=\frac{1}{\gamma_t^2}\frac{1}{|\eta|}\frac{p}{2.355\ \sigma _p}. \label{eq3}
\end{equation}
Therefore, the mass resolving power for each ion species depends critically on the phase-slip factor $\eta$ and the relative momentum difference $\frac{\sigma _p}{p}$ of the stored ions. The larger $|\eta|$ is, the poorer mass resolving power will be. In our recent IMS experiments conducted at CSRe, the momentum acceptance of the storage ring CSRe was about $\pm 2\times 10^{-3}$ \cite{ZHANGPRL_12}. In consequence, the so-called isochronous window, where isochronous conditions were best fulfilled, namely a relatively small $\eta$, was defined in the revolution time spectrum. Only nuclides within that window were used in the further mass determination.

In order to improve the mass resolving power and to expand the usable range of the revolution time spectrum, several approaches were explored. The first is the ${B\rho}$-tagging method which was realized at the FRS-ESR facility at GSI \cite{GeisselIMS,BHNPA1,BHNPA2}. During that experiment, the momentum acceptance was limited to $\frac{\Delta p}{p}=5\times 10^{-5}$ by utilizing a pair of slits at the second dispersive focal plane of the fragment separator FRS. The shortcoming of this method is that the transmission efficiency of the secondary beam is greatly reduced due to smaller acceptance through the fragment separator. So this method is not suitable for the mass measurements of ions with extremely low yields.

Another technique is being developed at the storage ring RI-RING which is under construction in RIKEN \cite{RIRING}. The velocity of each ion is measured in the beam line \cite{RIRING}, before injection into the ring. The effect of non-isochronicity can be corrected with velocities of ions \cite{RIRING}.

Another idea has been proposed to measure the velocities of stored ions in the ring \cite{GeisselIMS,DTOFP}, employing two TOF detectors installed in a straight section of the ring. On the basis of this idea, we have developed an upgraded IMS technique with two TOF detectors. The revolution times of the stored ions can be corrected properly via the additional velocity information obtained from the two TOF-detectors. Thus, the mass resolving power can significantly be improved while there is no need to limit the transmission efficiency of the secondary beam.

In this paper, we outline the principles of a new data analysis method describing how to apply the velocity information to the revolution times. The new method is tested with the aid of simulation. The motions of the stored ions were simulated with a newly developed software called SimCSRe \cite{SIM}.  Results obtained from the simulation data showed that the new IMS method can significantly
improve mass resolving power.

\section{Details of the two-TOF detector system}
\subsection{Scheme of IMS with two TOF detectors}
\begin{center}
\includegraphics[width=8.0cm]{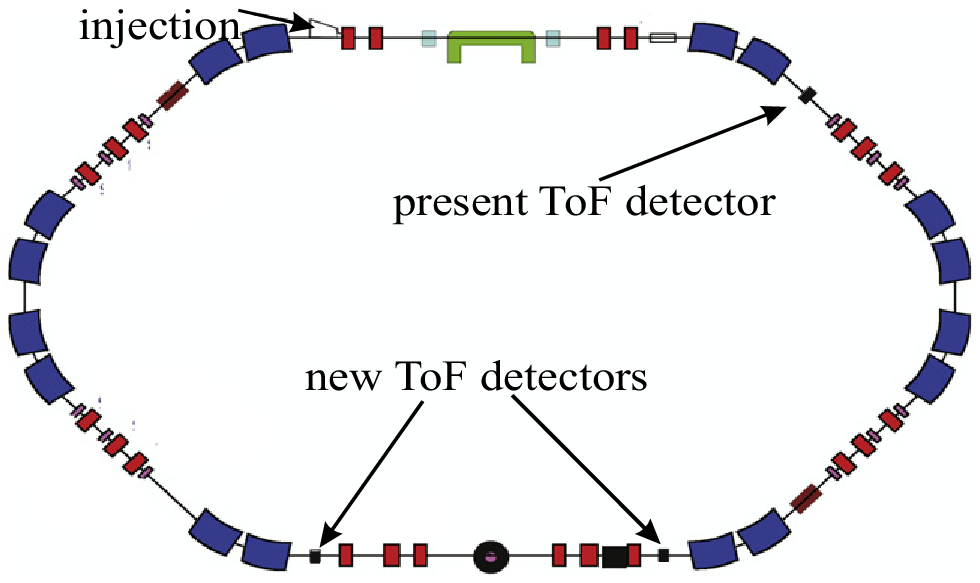}
\figcaption{\label{dtofcheme}Schematic view of the arrangement of  the two TOF detectors, which are installed in the straight section of the CSRe. }
\end{center}

In order to measure the velocities of stored ions, two TOF detectors with identical design were installed in the straight section of CSRe as shown in Fig.~\ref{dtofcheme} \cite{DTOF}. Between these two detectors, there is no dipole magnet and the distance is 17831 mm. Each detector consists of a carbon foil and a set of micro-channel-plates~(MCP) which collects secondary electrons released from the foil when stored ions pass through. More details of the detectors can be found in Refs \cite{MEIBO,ZW1,ZW2}.

\subsection{Principles of IMS with two TOF detectors.}

When an ion circulates in the storage ring, its closed orbit is determined by its magnetic rigidity. For a given ion species, the faster ions always circulate in the longer orbits while the slower ones move in the correspondingly shorter orbits. If the ring is set in the isochronous mode for this ion species, the difference of the orbit lengths shall compensate exactly the difference of velocities, making the revolution times independent of the velocities. In the IMS measurements, ions with a broad range of $m/q$ values are injected and stored in the ring simultaneously. These ions have very different velocities, while they circulate in similar orbits around the central orbit and within the acceptance of the ring. For an ion species that does not fulfill the isochronous condition, the revolution times are different for different orbits or different magnetic rigidities.

To describe the principles of the IMS with two TOF detectors, the revolution times of any given ion species are considered. Because there is no bending component in the straight section, the length of flight path between the two detectors is approximately the same for all ions. The velocity of one ion $v_i$ can be measured with the two TOF detectors, and is defined as follow:
\begin{equation}
  v_i=\frac{L}{t_{i,TOF1}-t_{i,TOF2}},\label{eq4}
\end{equation}
where $L$ is the path length of ions that fly over the two TOF detectors, and $t_{i,TOF1}$, $t_{i,TOF2}$ are the time stamps recorded by the two TOF detectors respectively.
The total orbit length $C_i$ of one ion can be determined to be:
\begin{equation}
C_i=T_i{\times}v_i,\label{eq5}
\end{equation}
where $T_i$ is the revolution time.

Let us define one reference orbit with length of $C_0$, and the velocity of that ion circulating in this orbit is $v_0$. The relationship between the orbits $C_0$ and $C_i$
is described as follows:
\begin{equation}
\frac{T_0-T_i}{T_i}=-(1-{\frac{\gamma _i ^2}{\gamma _t ^2}})\frac{v_0-v_i}{v_i}.\label{eq6}
\end{equation}
\begin{equation}
\frac{v_0-v_i}{v_i}=\frac{1}{\gamma _i ^2}\frac{B\rho _0 -B\rho _i}{B\rho _i}=\frac{\gamma _t ^2}{\gamma _i ^2}\frac{C_0-C_i}{C_i}.\label{eq7}
\end{equation}
Therefore, the revolution time $T_0$ of this ion in the reference orbit $C_0$ is:
\begin{equation}
T_0=\frac{C_0}{v_0}=T_i+(1-{\frac{\gamma_t^2}{\gamma_i^2}}){\frac{C_0-C_i}{C_i}}T_i \label{eq8}
\end{equation}

The revolution times of ions in any orbit can be converted to the revolution time corresponding to the same reference orbit $C_0$. In another word, the revolution times for ions with the same magnetic rigidity are obtained. In this way the time resolution, and consequently the mass resolution, for all ions can be improved.

In principle, the reference orbit $C_0$ can be defined arbitrarily, without influencing the final results. In the following analysis, the central orbit of CSRe is chosen as the reference.

In reality, an ion's orbit fluctuates around the closed orbit, with the amplitude of the fluctuation depending on the emittance. In consequence, the revolution times vary turn by turn. However, the revolution time over many turns is on average determined as the length of the closed orbit divided by its velocity. Since the energy losses of the ions in the TOF detectors are minor, the revolution times for different turns have little variation due to energy loss. In the measurements, the signals for one ion are recorded for around 300 revolution turns. In this analysis, the mean values of revolution time and velocity averaged over a total of 300 turns are used although in the real experiment, it is not feasible to obtain this kind of average because the typical detection efficiency is only $20-70 \%$. However, the standard fitting procedure which was applied in the data analysis for real experimental data \cite{TUNIMA} will obtain the same results.

\section{Monte-Carlo simulation of the IMS experiment}

\subsection{Principles of the simulation}
\begin{center}
\includegraphics[width=8.0cm]{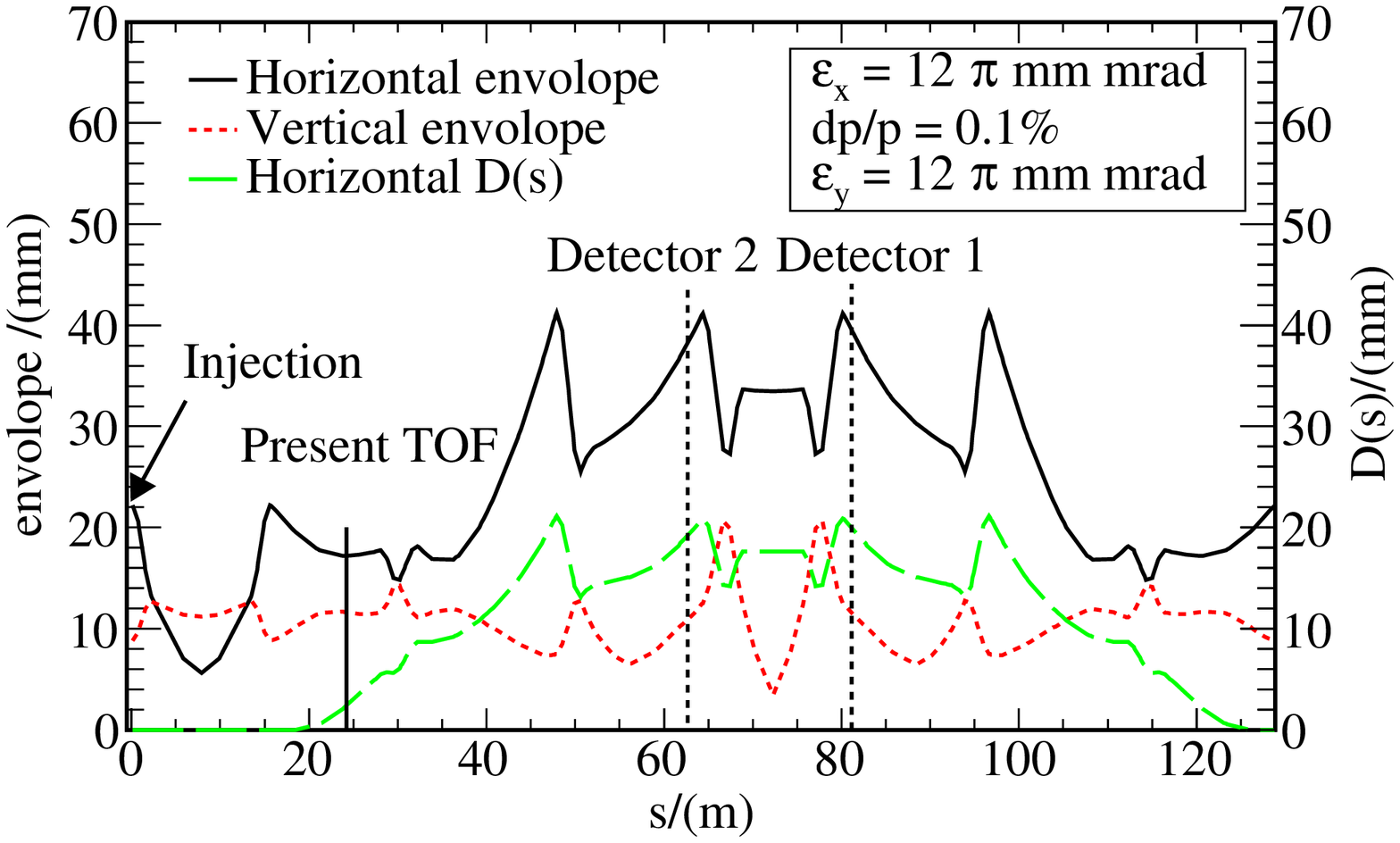}
\figcaption{\label{envolop} The simulated horizonal (thick black line) and vertical (red dotted line) beam envelopes and horizonal dispersion $D(s)$ (green dash line) as a function of the orbital length for an ion beam with $\varepsilon_{x,y} = 12$ $\pi$ mm mrad, $\frac{dp}{p}=10^{-3}$ . The positions where the present TOF detector and two new TOF detectors are installed are indicated. }
\end{center}

A simulation was conducted to test the principles of the upgraded IMS using two TOF detectors. Six-dimensional phase-space linear transmission theory was employed to simulate the motion of stored ions in the experimental storage ring CSRe. CSRe is composed of 144 elements and each element is represented by a 6-by-6-dimension first-order transfer matrix $M$. The basic algorithm is $B_f$ = $MB_i$, where $B_i$  and $B_f$  are six dimensional phase-space vectors of ions at the entrance and exit of each element of the CSRe lattice, respectively. The betatron motion of ions in the CSRe is simulated. The beam emittance, thickness of carbon foil, and size of carbon foil are also included in the simulation. The details of simulations can be found in Ref \cite{SIM}.  The simulation reproduces the experimental results very well.

Fig.~\ref{envolop} displays beam envelopes of the present isochronous mode of CSRe \cite{Xia_NIMA2002_488} calculated from the simulation program. As can be seen from the figure, the horizontal envelope at the position where the two TOF detectors are installed is very large.

In the simulation $\gamma_t$ is not an explicit parameter, while its value can be calculated according to the definition. The $\gamma_t$ value is shown as a function of the magnetic rigidity in Fig.~\ref{gtr}. The distribution can be fine-tuned with higher-order optical components~\cite{IC1,IC2}, which are not included in the simulation. We can see in the discussion that the inhomogeneous distribution of $\gamma_t$ dose not have much influence on the results.

\begin{center}
\includegraphics[angle=0,width=8cm]{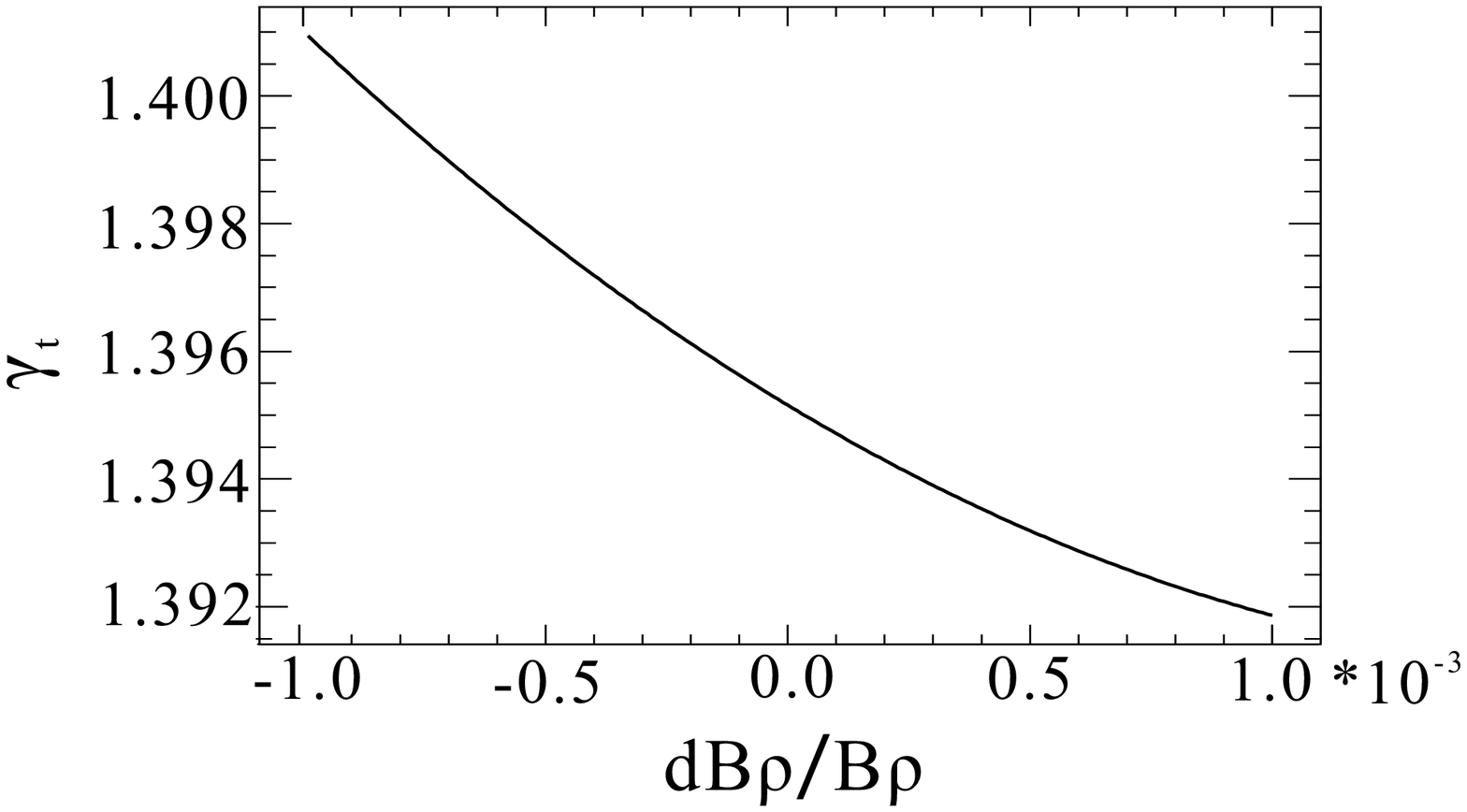}
\figcaption{\label{gtr}Simulated relationship between transition point $\gamma _t$ and momentum for a give ion species.}
\end{center}

In our simulation, each ion circulates 300 turns and loses energy in both TOF detectors when passing through their foils. The energy loss of each ion is calculated using the program LISE++ \cite{LISE} applying ATIMA program including L-S theory. The detailed description of that program can be found in Ref \cite{ATIMA}. For simplicity, the energy loss of a certain ion species at every revolution of any injection is assumed to be constant. The intrinsic time resolution of both TOF detectors is set to be 30 ps, comparable to the experimental value \cite{ZW1,ZW2}.

The simulated data were analyzed in two steps. Firstly, time information from only one TOF detector was analysed, which can be regarded as the traditional one-TOF IMS. The revolution times were extracted from the simulated data using the same data-analysis procedure as in the real experiments \cite{TUNIMA}. Secondly, the extracted revolution times of the stored ions were corrected by using the velocities obtained from two TOF detectors.

\subsection{Details of the data analysis}

Due to the relatively large dispersion at the positions where our two time-of-flight detectors were installed, we did two simulations. One is for TOF detectors with a carbon foil of 40 mm in diameter, which is the actual size of our present TOF detector. The other is for a virtual detector with a carbon foil of 100 mm. Under this condition, all ions within B$\rho$ acceptance $\pm 0.2\%$ can be stored in the ring, which is comparable to the practical $B\rho$ acceptance of our former IMS experiment \cite{ZHANGPRL_12}. In the simulation, the setting of the ring is the same as in the real experiment with $^{58}$Ni as the primary beam. The $\gamma _t$ of CSRe is set to be 1.395 and the mean $B\rho$ is set to be 5.6770 Tm.
\begin{center}
\includegraphics[angle=0,width=8.0cm]{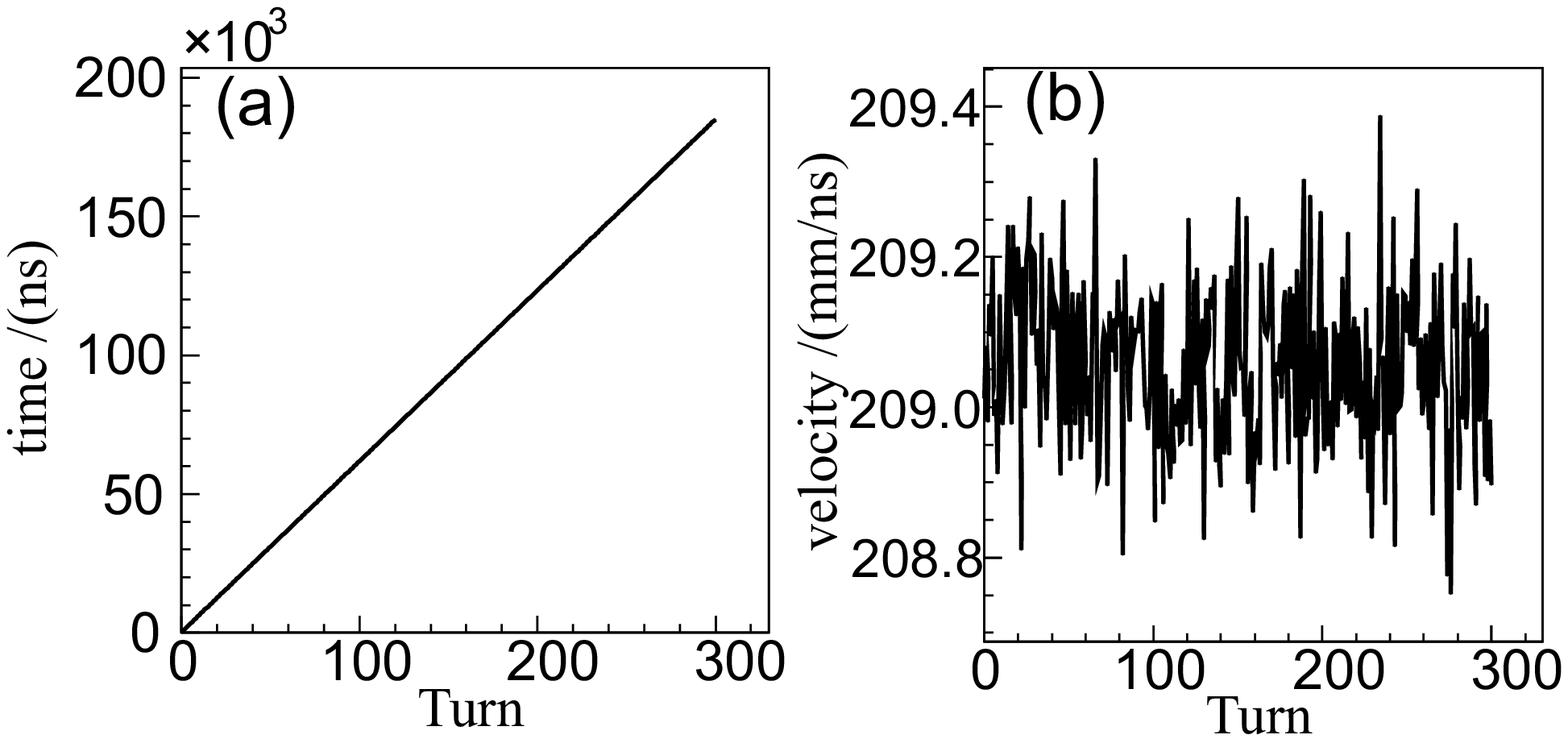}
\figcaption{\label{velocity}(a) Simulated time stamps from one TOF detector versus the corresponding revolution number for an $^{15}$O$^{8+}$ ion. (b) Velocities versus the corresponding revolution number for the $^{15}$O$^{8+}$ ion.}
\end{center}

From our simulation, the time stamps of every ion passing through both TOF detectors can be obtained. Fig.~\ref{velocity} (a) displays the time stamps from one TOF detector versus the corresponding revolution number for an $^{15}$O$^{8+}$ ion in a single injection. The obtained time stamps are fitted with a second order polynomial function of the revolution number. The revolution time of each ion is extracted from the slope of the fitting function at the $35^{th}$ revolution \cite{TUXLPRL_11}.
\begin{center}
\includegraphics[angle=0,width=8cm]{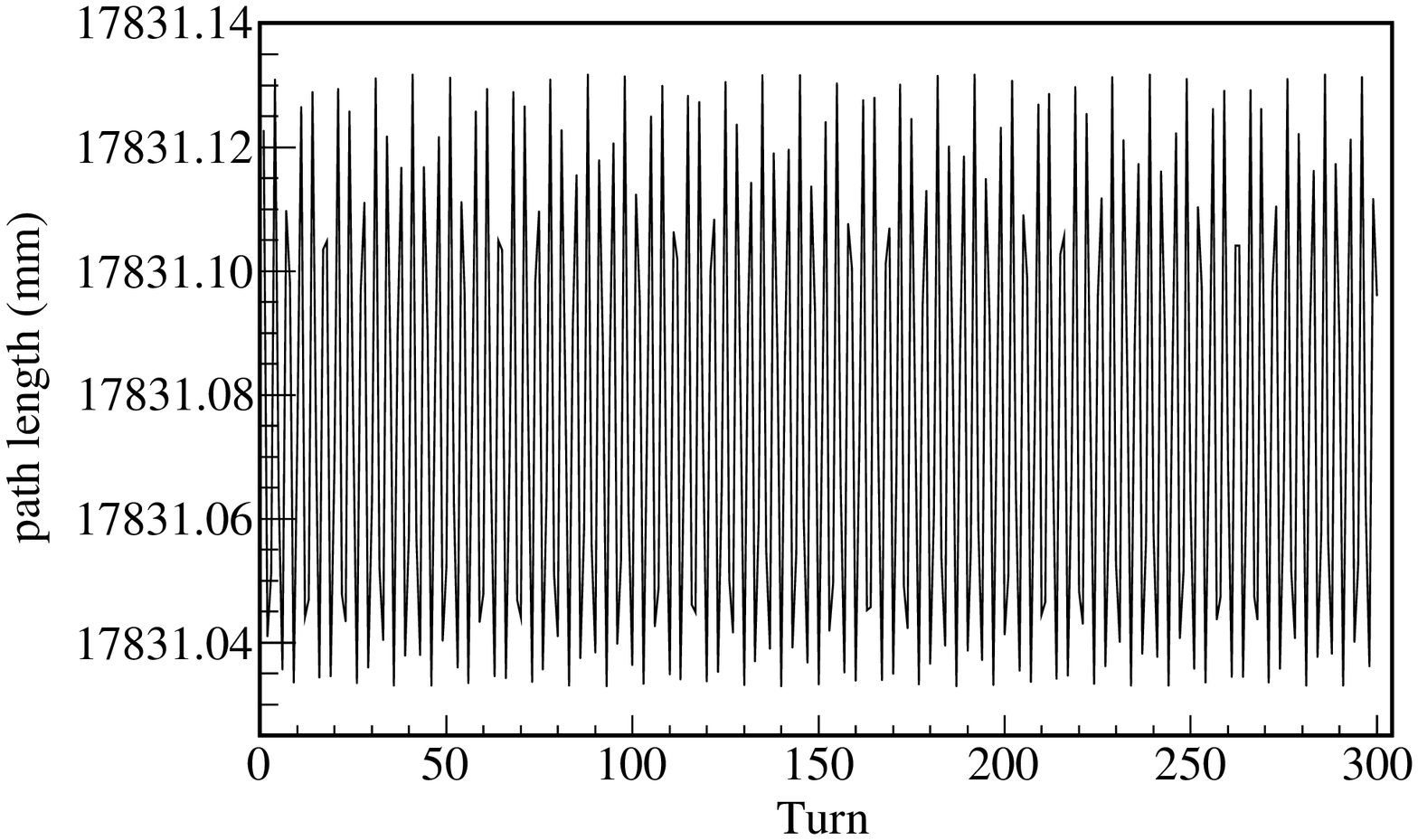}
\figcaption{\label{path}Path length of the $^{15}$O$^{8+}$ ion flight over the straight section at each revolution obtained from the simulation. The
diameter of the carbon foils in this simulation were 100 mm.}
\end{center}

The velocity is calculated by using Eq.~(\ref{eq4}). The flight time between two TOF detectors was obtained from revolution time stamps of both TOF detectors at the same turn. In real experiments as well as in our simulations, the path lengths of the stored ions that fly over the straight section change turn-by-turn due to the betatron oscillation. Fig.~\ref{path} shows the path lengths of this $^{15}$O$^{8+}$ ion passing over the two TOF detectors at each revolution. The path length can not be known in the real experiments. In the data analysis for simulations, the path lengths are all assumed to be equal to the geometrical distance between the two TOF detectors of 17831 mm. In the IMS using two TOF detectors with carbon foils of 100 mm, the maximum relative error of the velocity cause by this assumption for the path length is about $\Delta L/L$ = $1.6 \times 10^{-5}$. The intrinsic time resolution of both TOF detectors is 30 ps, and the flight time between the two TOF detectors is about 85 ns. The relative error caused by the TOF detectors is $\sigma_t/t$ = $30 \times \sqrt{2} \times 10^{-3}/85$ = $5.0 \times 10^{-4}$. This is at least 30 times larger than $\Delta L/L$.

Fig.~\ref{velocity} (b) displays calculated velocities versus the corresponding revolution number for the $^{15}$O$^{8+}$ ion. It is clear that the calculated velocities fluctuate randomly for different revolution number. The relative difference between the maximum and the minimum velocities is about $\Delta v/v$ $\approx$ $2.8 \times 10^{-3}$ and the relative error of velocity for the $^{15}$O$^{8+}$ ion determined in this method is $ \sigma_ {v} /{v} = 5.2 \times 10^{-4}$. This number is consistent with $\sigma_t/t$ as mentioned above. We can conclude that the limitation to velocity determination is mainly due to the performance of TOF detectors in our simulations. In other words, the assumption for the path length is reasonable.

For every ion stored in the CSRe, its average revolution time and average velocity over 300 revolutions can be extracted. Its average orbital length can be calculated using Eq.~(\ref{eq5}). Fig.~\ref{fig04} (a) and Fig.~\ref{fig05} (a) display scatter plots of the average revolution time and  average orbital length for all $^{15}$O$^{8+}$ (whose $\gamma $ is close to $\gamma_t$) and $^{25}$Al$^{13+}$ (whose $\gamma $ is much smaller than $\gamma_t$) ions, respectively.
\begin{center}
\includegraphics[angle=0,width=8.0cm]{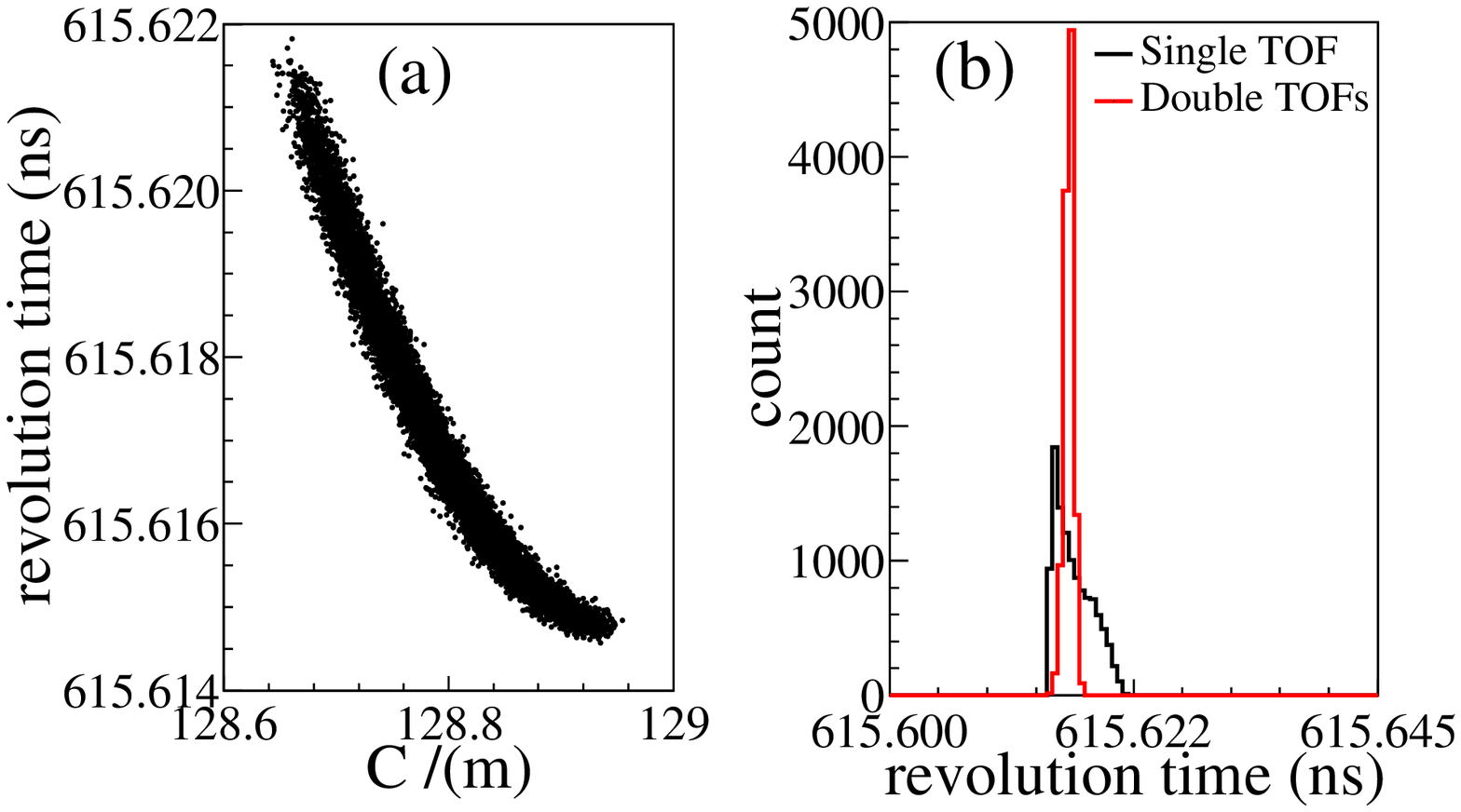}
\figcaption{\label{fig04}(a) Scatter plot of average revolution times versus average orbital length for all $^{15}$O$^{8+}$ ions.
(b) Spectra of the original and corrected revolution time for all $^{15}$O$^{8+}$ ions.
The diameter of the carbon foils here were 100 mm.}
\end{center}
\begin{center}
\includegraphics[angle=0,width=8.0cm]{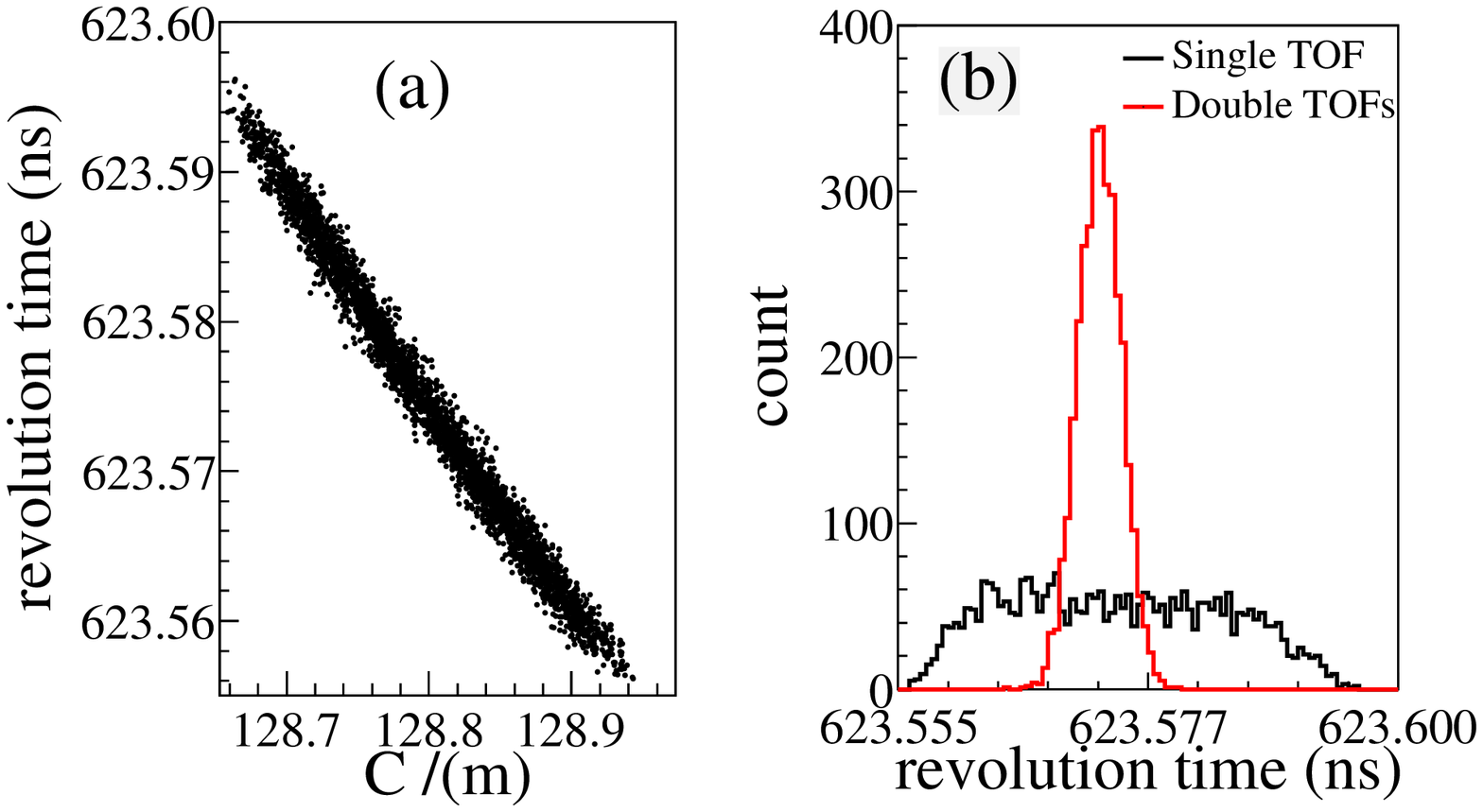}
\figcaption{\label{fig05}Same as Fig.~\ref{fig04} but for all $^{25}$Al$^{13+}$ ions.}
\end{center}
The revolution times can be corrected to that corresponding to the central orbit as described in Eq.~(\ref{eq8}). The transition point $\gamma _t$ is set to be 1.395.
Fig.~\ref{fig04} (b) and Fig.~\ref{fig05} (b) show the comparison between the original average revolution time spectrum obtained from one TOF detector and the corrected revolution time spectrum. It is clear that the resolving power of revolution time is improved by 5.0 times for the $^{15}$O$^{8+}$ ions and 5.8 times for the $^{25}$Al$^{13+}$ ions .

We should note that the actual size of carbon foil of the present TOF detector is only 40 mm in diameter, which will limit the momentum acceptance if the two TOF detectors are installed at the position as indicated in Fig.~\ref{dtofcheme}. Therefore, a larger size TOF detector is under development. Fig.~\ref{fig06} displays the comparison between the original average revolution time spectrum and the corrected revolution time spectrum for all ions species from simulations with different TOF parameters.
\begin{center}
 \centering
\includegraphics[angle=0,width=8.0cm]{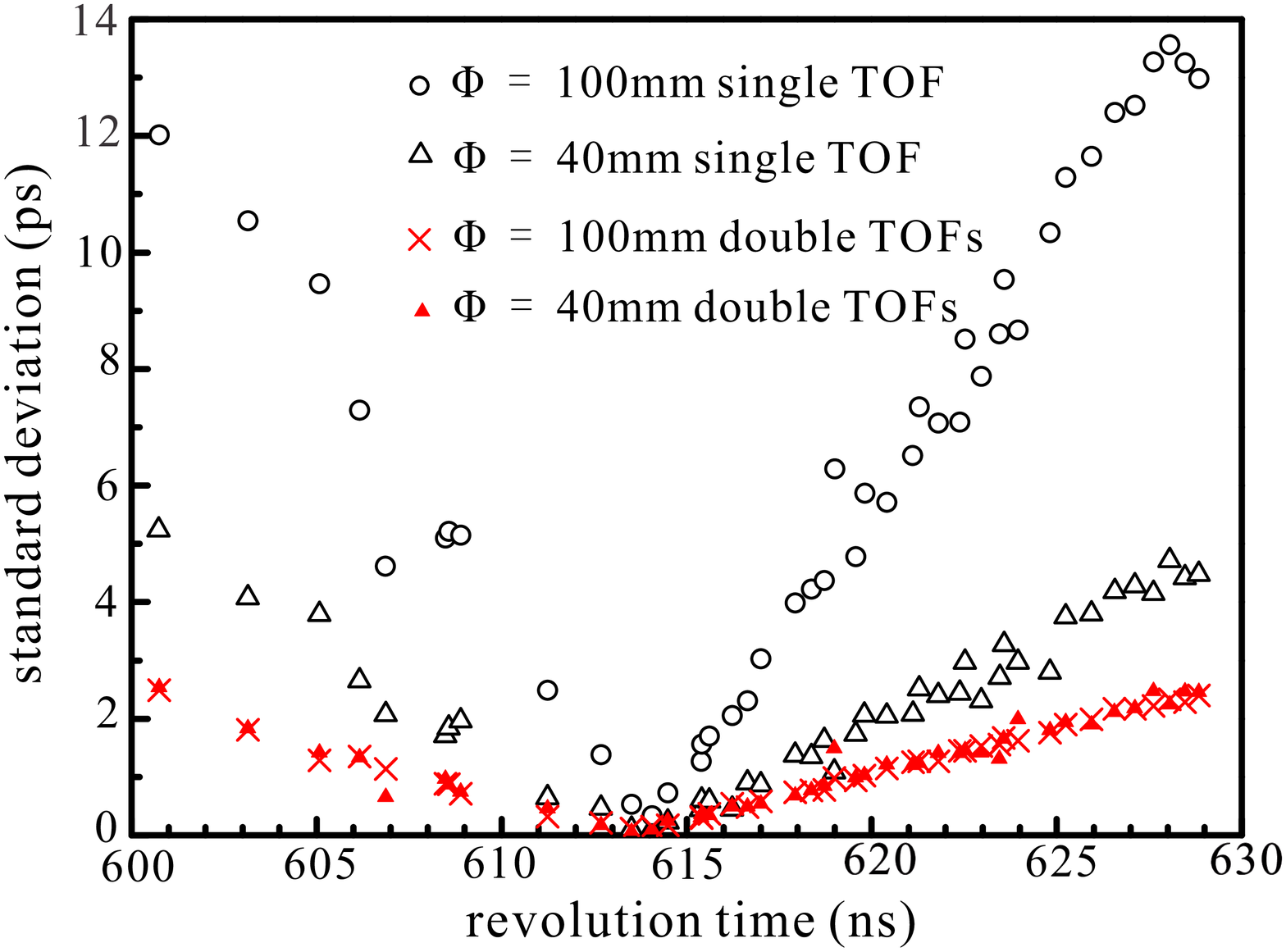}
\figcaption{ \label{fig06} Standard deviations of the revolution time for each nuclide obtained from simulations. The red solid triangles and black open triangles represent the simulation results analyzed using time stamps from two TOF detectors and single TOF detector respectively assuming the size of the carbon foil is 40 mm.  The red crosses and open circles show the simulation results assuming that the size of the carbon foil is 100 mm.}
\end{center}

 As can be seen from Fig.~\ref{fig06}, smaller standard deviation $\sigma _T$ for all nuclides is achieved with one TOF detector with $\Phi =$ 40 mm (black open triangles) compared to  that with $\Phi =$ 100 mm (black open circles). This is due to the smaller size of TOF detectors resulting in the reduction of momentum acceptance in the CSRe. This effect is similar to the effect of the $B\rho$-tagging method before injection into the storage ring \cite{GeisselIMS,BHNPA1,BHNPA2}, which is at the cost of largely reduced acceptance.

 The $\sigma _T$ of all nuclides obtained from two TOF detectors with $\Phi =$ 100 mm (red crosses) are reduced by about 5 times while those from two TOF detectors with $\Phi =$ 40 mm (red solid triangles) are reduced by about 2 times compared to the respective results for a single TOF.
 However, the $\sigma _T$ of all nuclides obtained from two TOF detectors are almost the same, even though the momentum acceptance in the former simulation is much larger than in the latter. Within the same beam time, a larger momentum acceptance in the CSRe will lead to higher statistics. Therefore, in order to improve the performance of IMS with two TOF detectors, a larger size TOF detector or a new lattice with smaller beam envelope at the straight section, which will result in a large momentum acceptance, is very beneficial and thus necessary.

 With this new IMS, more distinct improvement of nuclides with $\gamma$ far away from $\gamma_t$ was achieved compared to nuclides with $\gamma$ closed to $\gamma_t$. It means that the upgraded IMS can significantly enlarge the isochronous window and more nuclei can be used in the mass analysis. This is very useful for the measurements for nuclei far away from $\beta$-stability.
 \begin{center}
 \centering
\includegraphics[angle=0,width=8.0cm]{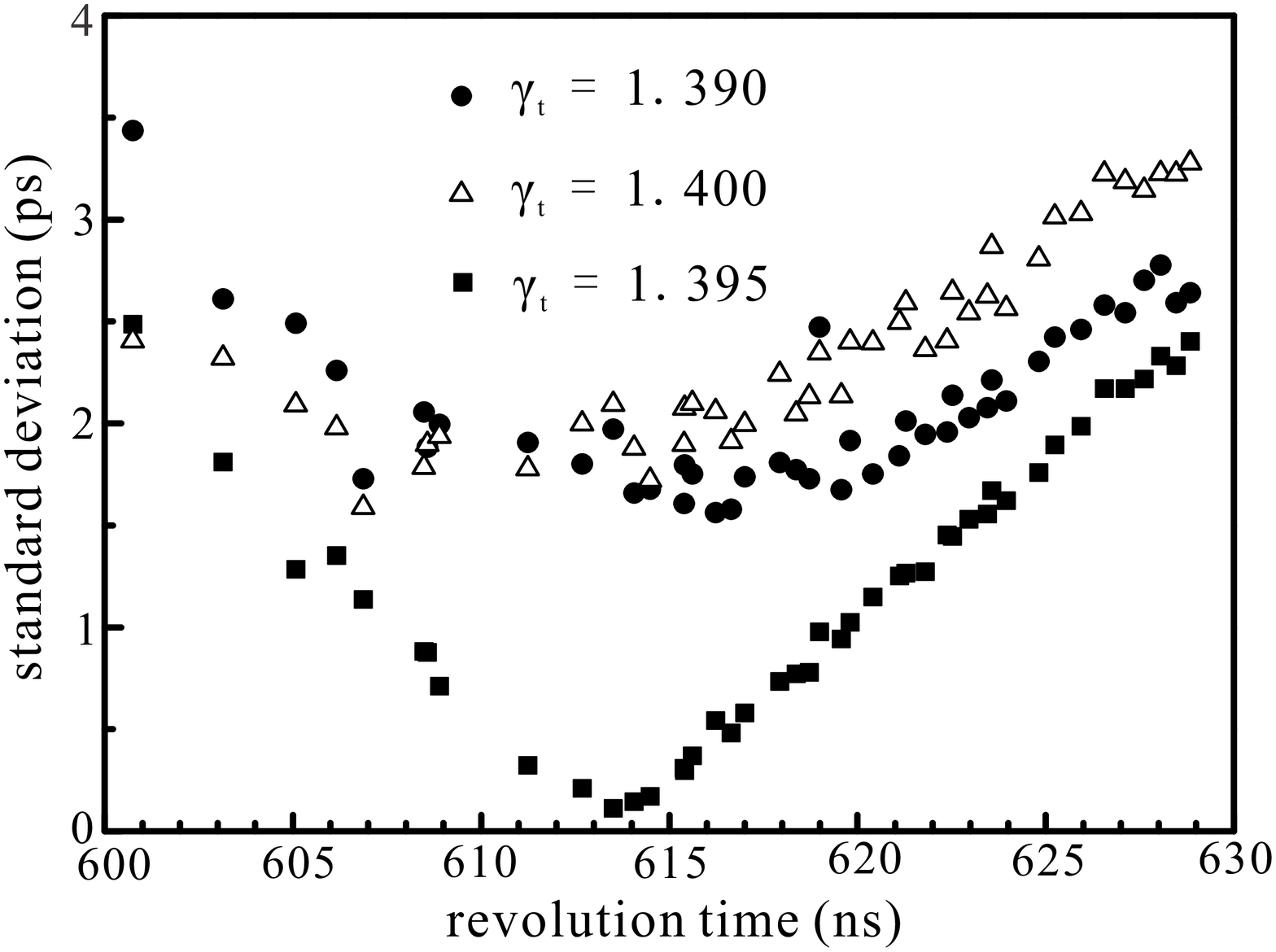}
\figcaption{ \label{gammat} Results from the same simulated data but applying different $\gamma_t$ values than 1.395 in Eq.~\ref{eq8}. Solid circles, open triangles and solid squares represent results for $\gamma_t$ = 1.390, 1.400, and 1.395,  respectively. This simulation is for the IMS with two TOF detectors of $\Phi$ =100 mm.}
\end{center}

 In the former data analysis, an important parameter $\gamma_t$ in Eq.~\ref{eq8} was set to be 1.395. In reality, $\gamma_t$ can not be known accurately in experiments. In order to explore the effect of the unknown $\gamma_t$ in real experimental data analysis, $\gamma_t$ was set to be 1.390, 1.395 and 1.400 in three data analyse for the same data. Fig.~\ref{gammat} presents the comparison of the results. As can be seen from that figure, if $\gamma_t$ is not properly set, the systematic behavior of $\sigma_T$ with the revolution time will be distorted. Taking this as a criterion, $\gamma_t$ can be scanned in the data analysis for the experimental data and a proper $\gamma_t$ can be obtained within the relative uncertainty of $3.6 \times 10^{-3}$.

 As discussed before, the precision of velocity determination is mainly limited by the performance of the TOF detectors. According to Eq.~\ref{eq8}, the accuracy of the correction for revolution times
depends on the accuracy of the velocity determination. As a consequence, the mass resolving power of the IMS with two TOF detectors is limited by the performance of the TOF detectors. Fig.~\ref{resolution} illustrates the effect of intrinsic time resolution of the two TOF detectors on the mass resolving power. In the ideal case of 0 ps intrinsic time resolution, $\sigma_T$ is close to zero, and is almost the same for all ion species.
 \begin{center}
 \centering
\includegraphics[angle=0,width=8.0cm]{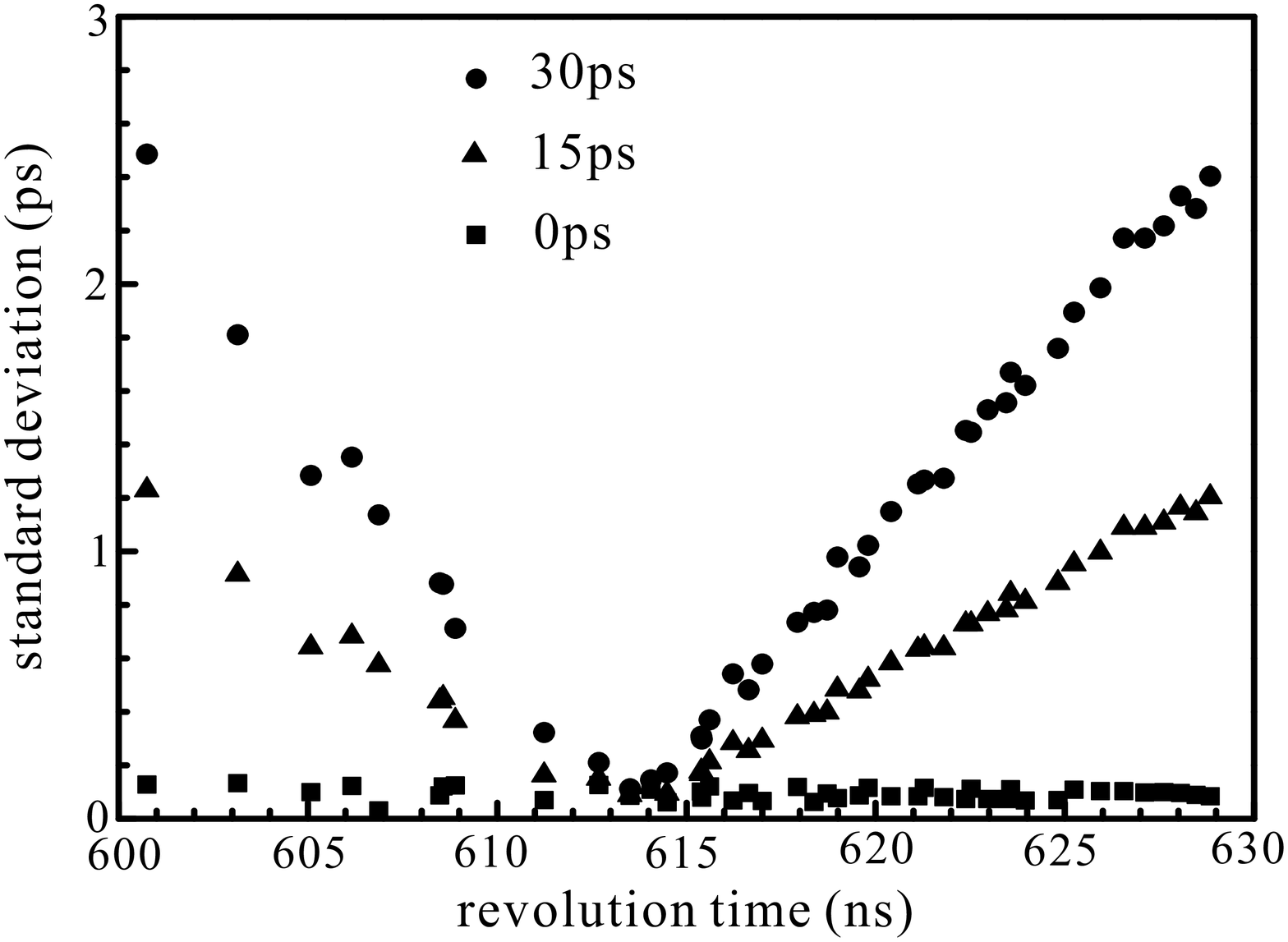}
\figcaption{ \label{resolution} Results from three simulations for IMS with two TOF detectors by changing the  intrinsic
time resolution of both TOF detectors to 30 ps, 15ps, and 0ps. The diameter of the carbon foils  were 100 mm. }
\end{center}

\section{Summary}
We have developed a new methodology for the IMS with two TOF detectors. The new method was applied to a series of simulated data. The simulated data were analyzed and the velocities of the circulating ions were extracted using the new approach. The revolution time of each ion was then corrected to the one corresponding to the reference orbit, employing the additional velocity information. The results showed that the IMS with two TOF detectors can improve the mass resolving power without limiting the acceptance of the ring. Compared to the IMS with only one TOF detector, the mass resolving power of the upgraded IMS with two TOF detectors can be significantly improved, especially for the nuclei with $\gamma$ far away from $\gamma_t$. The mass resolving power of the IMS with two TOF detectors is limited predominantly by the performance of the two TOF detectors.


\end{multicols}

\vspace{15mm}

\vspace{-1mm}
\centerline{\rule{80mm}{0.1pt}}
\vspace{2mm}

\begin{multicols}{2}

\end{multicols}

\clearpage

\end{document}